\documentclass[aps,prl,twocolumn]{revtex4}
\usepackage{epsfig}
\newcommand{\msbar}{\overline{\mbox{MS}}}
\newcommand{\eps}{\varepsilon}
\begin{document}
\preprint{MZ-TH/06-06}
\title{Large next-to-leading order QCD corrections to pentaquark sum rules}
\author{S.~Groote,$^{1,2}$ J.G.~K\"orner$^{2,3}$ and
  A.A.~Pivovarov$^{2,4}$}
\affiliation{$^1$Tartu \"Ulikooli Teoreetilise F\"u\"usika Instituut,
  T\"ahe 4, EE-51010 Tartu, Estonia\\
  $^2$ Institut f\"ur Physik, Johannes-Gutenberg-Universit\"at,
  Staudinger Weg 7, D-55099 Mainz, Germany\\
  $^3$ National Centre of Physics, Quaid-i-Azam University Campus,
  Islamabad, Pakistan\\
  $^4$ Institute for Nuclear Research of the
  Russian Academy of Sciences, Moscow 117312}

\begin{abstract}\noindent
Using configuration space techniques, we calculate next-to-leading order (NLO)
five-loop QCD corrections to the correlators of interpolating pentaquark
currents in the limit of massless quarks. We obtain very large NLO corrections
to the spectral density which makes a standard sum rule analysis problematic.
However, the NLO corrections to the correlator in configuration space are
reasonable. We discuss the implications of our results for the
phenomenological sum rule analysis of pentaquark states.
\end{abstract}

\maketitle

\section{Introduction}
There is a continuing interest in exotic states of strong interactions that
differ from the standard nonexotic mesons and baryons (see
e.g.~\cite{Matveev,Braun:1988kv}). Nonexotic three-quark baryons have been
intensively studied in ~\cite{Ioffe,Chung}. Finite mass effects were studied 
in~\cite{Pivovarov:1988gt,Ovchinnikov,Groote} where the next-to-leading order
(NLO) QCD corrections to the correlators of finite mass baryonic currents were
determined. This led to an improved precision of the sum rule analysis.
Different aspects of the physics of multiquark states in QCD have been
discussed long ago~\cite{Larin:1985yt}. It is noteworthy that the analysis of
multiquark exotics can provide valuable information on the details of
quark-gluon interaction relevant to nuclear
physics~\cite{Meissner:2004yy,Bijnens:2005mi}.

Candidates to be described in QCD as multiquark states are the
deuteron~\cite{LarMat,Kulagin:kb} or Jaffe's dihyperon
$H$~\cite{Jaffe,Sakai:1999qm}. Since the dihyperon state can only decay 
weakly, it would be expected to be quite narrow. To the best of our knowledge
the dihyperon state is the first state to attract attention in the modern
context of QCD. The discovery of gluon bound states would be a triumphant
confirmation of QCD and would allow for a quantitative check of QCD in the
completely new sector of the glueball states~\cite{Brodsky:2003hv,%
Kataev:1981aw}. Another class of multiquark states, the pentaquark states,
have recently become the focus of intense theoretical and experimental studies.

The properties of all of these multiquark (or multigluon) states can be 
studied in a model independent way through the method of the QCD sum rule
analysis. In the QCD sum rule analysis one analyzes the operator product
expansion of current--current correlators of interpolating local fields which
have the quantum numbers of the multiquark states under study. It is well
known that radiative corrections have a strong impact on the results of the
sum rule analysis. In this paper we derive the necessary tools that allow
one to compute the $\alpha_S$ radiative corrections to multiquark sum rules
in the limit when the quarks (or antiquarks) are massless. As a specific
example we apply our method to pentaquark correlators and calculate the NLO
radiative corrections to the pentaquark current correlator and the spectral
density of a specific pentaquark current.

\section{Calculation}
An important ingredient in the formulation of the operator product expansion
(OPE) analysis for the pentaquark states is the choice of the interpolating
current. The result depends strongly on the choice of the interpolating
current as has been pointed out before in the sum rule analysis of the
dibaryon~\cite{Larin:1985yt}. The same holds true for the sum rule analysis of
pentaquark states~\cite{Matheus:2004qq,Matheus:2004gx,Lee:2005ny}. A detailed
analysis of the dependence on the interpolating current is out of the scope of
a short letter, it will be published elsewhere~\cite{Groote:2006}. In the
following we shall present the tools needed to calculate the NLO QCD
corrections for interpolating quark currents of any composition. 

\subsection{Generalities}
In the QCD sum rule analysis the prime object of study is the correlation 
function
\begin{equation}
\Pi(q)=i\int d^4xe^{iqx}\langle 0|Tj(x)\bar j(0)|0\rangle,
\end{equation}
where the interpolating current $j(x)$ is a local operator with the quantum 
numbers of the pentaquark baryon state $\Theta$. It has to be constructed 
from four
quark and one antiquark fields such that its projection onto the pentaquark
state $|\Theta(p)\rangle$ is nonzero:
\begin{equation}
\langle 0|j(0)|\Theta(p)\rangle=\lambda_\Theta,\quad p^2=m_\theta^2.
\end{equation}
Since the interpolating current $j(x)$ is not unique, one immediately faces
the question of the optimal choice for the interpolating currents. Recall 
that the problem of choosing the optimal current already arose in the case of
baryons~\cite{Ioffe,Chung} where the currents are constructed from three quark
fields. In the pentaquark case there are five quark (antiquark) fields to 
build the interpolating currents and correspondingly the number of independent
currents with the correct quantum numbers is much larger
(see~\cite{Matheus:2004gx} and references therein). 

The treatment of the interpolating current $j(x)$ in its most general form, 
i.e.\ in the form of linear combinations of all possible independent local 
operators with the correct pentaquark quantum numbers, is a very unwieldy 
problem. In this pilot study of the NLO corrections to interpolating 
pentaquark currents we limit ourselves to the study of only one simple
interpolating currents with the required pentaquark quantum numbers. As in the
case of mesons and baryons, we construct our interpolating current $j(x)$ from
quark fields without derivatives.

\subsection{Tools for the NLO calculation}
\begin{figure}[htb]\begin{center}
\includegraphics[angle=0,width=0.15\textwidth]{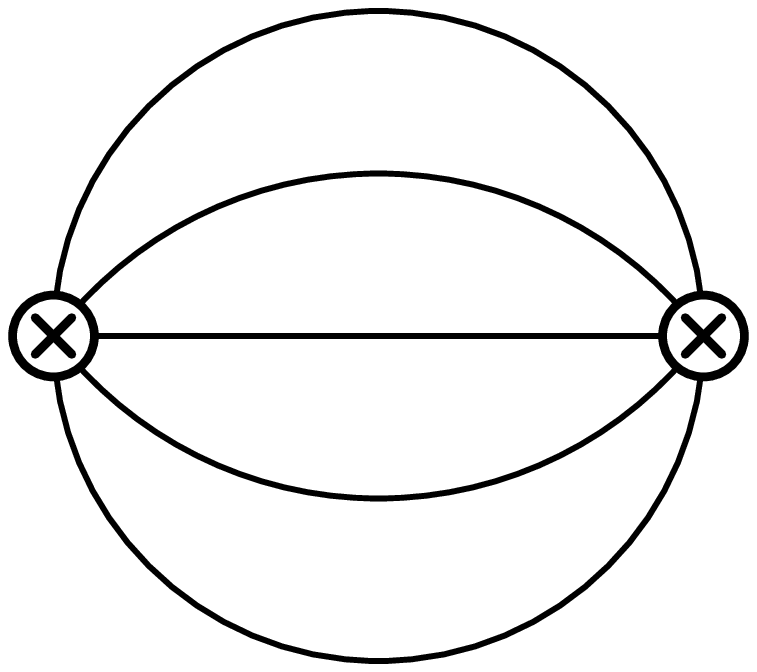}
\includegraphics[angle=0,width=0.15\textwidth]{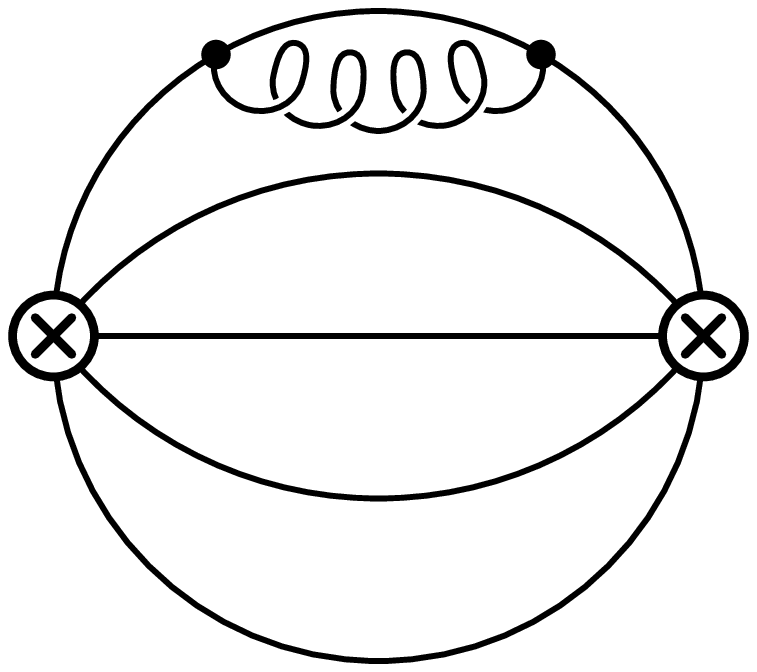}
\lower4pt\hbox{\includegraphics[angle=0,width=0.15\textwidth]{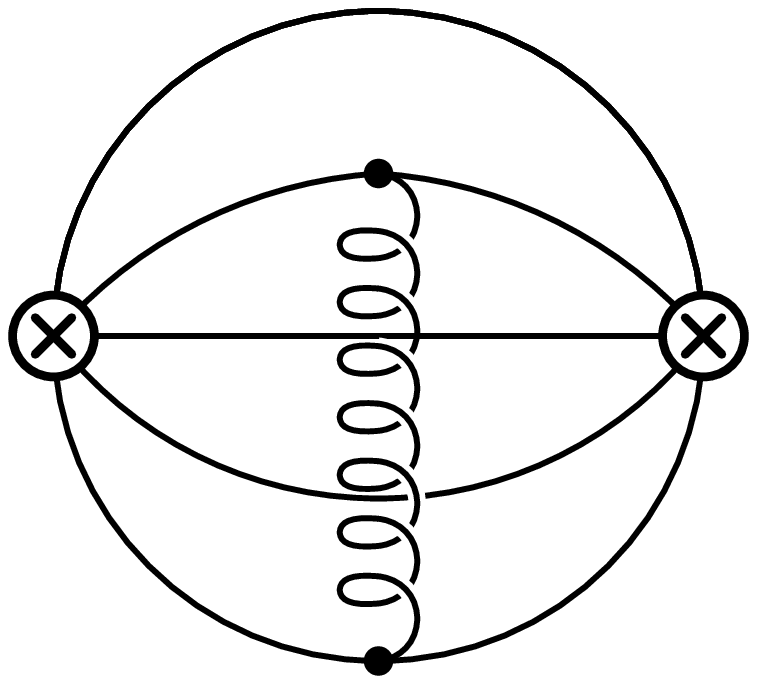}}
\vspace{7pt}
\centerline{(a)\kern68pt(b)\kern68pt(c)}
\caption{\label{pentcor}LO contribution (a) and examples for the NLO propagator
 (b) and dipropagator corrections (c)}
\end{center}\end{figure}

The LO diagram for the pentaquark correlator is given by the four-loop
diagram FIG.~\ref{pentcor}(a).
There are two types of NLO five-loop corrections. First there are the one-loop
corrections to single quark propagators $S(x)$, one of which is shown in 
FIG.~\ref{pentcor}(b). We shall refer to these corrections as the propagator
corrections. Then there are the dipropagator corrections connecting two
different quark propagators, one of which is shown in FIG.~\ref{pentcor}(c). 

It turns out that it is very convenient to calculate these corrections in 
configuration space, in particular if the quarks are treated as 
massless~\cite{Groote:2005ay}. For the propagator correction $S_1(x)$ we obtain
\begin{eqnarray}\label{propagator}
S_1(x)|_{\rm NLO}&=&S_1(x)|_{\rm LO}
\left\{1-C_F\frac{\alpha_s}{4\pi}\frac1\eps\left(\mu^2_Xx^2\right)^\eps
  \right\}\nonumber\\
  &=&S_0(x^2)\gamma^\mu x_\mu\left\{1-C_F\frac{\alpha_s}{4\pi}\frac1\eps
  \left(\mu^2_Xx^2\right)^\eps\right\}
\end{eqnarray}
where in the Euclidean domain one has
\begin{equation}
S_0(x^2)=\frac{-i\Gamma(2-\eps)}{2\pi^{2-\eps}(x^2)^{2-\eps}}.
\end{equation}
$\mu_X$ is a renormalization scale in dimensional regularization appropriate
for calculations in configuration space. The space-time dimension is
parametrized by $D=4-2\eps$. This choice of renormalization scale avoids the
appearance of $\ln(4\pi)$ and $\gamma_E$ factors in configuration space
calculations. The relation of $\mu_X$ and the usual renormalization scale
$\mu$ of the $\msbar$-scheme is given by
\[
\mu_X=\mu e^{\gamma_E}/2.
\]
The dipropagator two-loop amplitude for a pair of quarks with open Dirac
indices leads to an integral that is a bit more difficult to calculate. The
result for the dipropagator correction including the LO term reads 
\begin{eqnarray}\label{dipropagator}
\lefteqn{S_2(x)|_{\rm NLO}
  \ =\ S_0(x^2)^2\{\gamma^\mu x_\mu \otimes\gamma^\nu x_\nu}\nonumber\\&&\strut
  +t^a\otimes t^a\frac{\alpha_s}{4\pi}\frac1\eps(\mu_X^2 x^2)^\eps
  (\gamma^\mu\otimes\gamma^\nu(a_1x_\mu x_\nu+b_1x^2g_{\mu\nu})
  \nonumber\\&&\strut\qquad\qquad\qquad\qquad
  +a_3\Gamma_3^{\alpha\beta\mu}\otimes{\Gamma_{3\ \alpha\beta}}^\nu
  x_\mu x_\nu)\}
\end{eqnarray}
where the coefficients $a_1$, $b_1$ and $a_3$ are given by 
\[
a_1=-1-\frac{11}2\eps,\quad
b_1=-1-\frac12\eps,\quad
a_3=-\frac12-\frac14\eps,
\]
and where
\begin{equation}
\Gamma_3^{\mu\alpha\nu}=\frac12(\gamma^\mu\gamma^\alpha\gamma^\nu-
\gamma^\nu\gamma^\alpha\gamma^\mu).
\end{equation}
Eqs.~(\ref{propagator}) and~(\ref{dipropagator}) allow one to calculate the
NLO corrections to $n$-quark(antiquark) current correlators of any composition
using purely algebraically algorithms without having to compute any integrals.

\subsection{Renormalization of the interpolating current}
In a NLO calculation one has to account for mixing effects between
operators when going through the renormalization program. Mixing can occur
when gluons are exchanged between the lines in the pentaquark correlator
(dipropagator corrections). 
In order to keep track of flavours we first treat the case of an 
interpolating current composed of five massless
quark(antiquark) fields with different flavours,
\begin{equation}
j=\eps^{ijk}(q_1^{iT}Cq_2^j)q_3^k(\bar q_4^lq_5^l)
\end{equation}
where $C$ is the charge conjugation matrix. 
The interpolating current consists of a baryonic part
$B=\eps^{ijk}(q_1^{iT}Cq_2^j)q_3^k$ and a mesonic part $M=(\bar q_4^lq_5^l)$.
If the gluon is exchanged within the mesonic part, the renormalization
factor is the usual one for the mesonic operator (see
e.g.~\cite{Pivovarov:1991nk}),
\begin{equation}
Z_M=1-\frac{\alpha_s}{\pi\eps}.
\end{equation}
If the gluon is exchanged within the baryonic part, the renormalization
factor is given by the known renormalization factor for the baryonic
operator~\cite{Pivovarov:1991nk},
\begin{equation}
Z_B=1-\frac{\alpha_s}{2\pi\eps}.
\end{equation}
These two contributions will be referred to as factorizing contributions. 
When the gluon is exchanged between the mesonic and baryonic part one 
obtains new operators which do not have the structure of the
original current. There are two contributions of this kind which we refer to
as mixed contributions. If the gluon is exchanged between the mesonic part and
the $(q_1^{iT}Cq_2^j)$ piece of the baryonic part, we obtain a contribution of
the operator
\begin{equation}
O_1=\eps^{ijk}\Big\{(q_1^{iT}C\sigma^{\alpha\beta}q_2^l)
  +(q_1^{lT}C\sigma^{\alpha\beta}q_2^i)\Big\}q_3^k
  (\bar q_4^l\sigma^{\alpha\beta}q_5^j)
\end{equation}
where $\sigma^{\alpha\beta}=\frac{i}2[\gamma^\alpha,\gamma^\beta]$. If the
gluon is exchanged between the mesonic part and the $q_3^k$ piece of the
baryonic part one obtains a contribution of the operator
\begin{equation}
O_2=\eps^{ijk}(q_1^{iT}Cq_2^j)\sigma^{\alpha\beta}
  \Big\{q_3^l(\bar q_4^l\sigma_{\alpha\beta}q_5^k)
  -\frac13q_3^k(\bar q_4^l\sigma_{\alpha\beta}q_5^l)\Big\}.
\end{equation}
The renormalized current reads
\begin{equation}
j_R^{\rm NLO}=Z_MZ_Bj^{\rm NLO}-\frac{\alpha_s}{16\pi\eps}\left(O_1+O_2\right).
\end{equation}

\section{NLO results}
In order to obtain NLO results, let us consider the interpolating current
\begin{equation}
j=\eps^{abc}(u^{aT}Cd\,^b)d\,^c(\bar s\,^e u^e)
\end{equation}
with the quantum numbers of the pentaquark baryon $\Theta$. The current is
constructed in such a way that it cannot directly dissociate into a neutron
and a kaon since the respective color singlet parts have the wrong parity.

The result of our NLO calculation for the bare correlator reads
$(\bar{j}=\eps^{abc}(\bar{d}\,^a C^{-1} \bar{u}^{bT}) \bar{d}\,^c 
(\bar{u}^e s^e))$
\begin{equation}
\langle 0|Tj(x)\bar j(0)|0\rangle=S_0(x^2)^5(x^2)^2\gamma^\mu x_\mu\Pi(x^2)
\end{equation}
where
\begin{eqnarray}\label{Pix2}
\Pi(x^2)&=&360\left(1+\frac{\alpha_s}\pi(\mu_X^2x^2)^\eps
  \left(\frac3\eps+3\right)\right) \nonumber\\&&
  -6\left(1+\frac{\alpha_s}\pi(\mu_X^2x^2)^\eps
  \left(\frac{11}\eps-\frac73\right)\right)\nonumber\\&&
  -4\cdot 6\left(1+\frac{\alpha_s}\pi(\mu_X^2x^2)^\eps
  \left(\frac9\eps-1\right)\right).
\end{eqnarray}
The first line contains the factorizing NLO contibutions while the second and
third lines contain the mixing contributions. The counter term contains
contributions from the renormalization factors $Z_M$ and $Z_B$ as well as from
the operators $O_1$ and $O_2$. It reads
\begin{equation}\label{DeltaPi}
\Delta\Pi=-\frac{\alpha_s}\pi\left\{360\frac3\eps-6\left(\frac{11}\eps
  -\frac{28}3\right)-4\cdot 6\left(\frac9\eps-7\right)\right\}.
\end{equation}
Adding Eqs.~(\ref{Pix2}) and~(\ref{DeltaPi}) one obtains the renormalized
correlator
\begin{eqnarray}\label{PiRx2}
\lefteqn{\Pi_R(x^2)\ =\ \Pi(x^2)+\Delta\Pi
  \ =\ 360\left(1+\frac{\alpha_s}{\pi}\left(3+3L_x\right)\right)}\nonumber\\&&
  -6\left(1+\frac{\alpha_s}{\pi}\left(7+11L_x\right)\right)
  -4\cdot 6\left(1+\frac{\alpha_s}{\pi}\left(6+9L_x\right)\right)\nonumber\\
\end{eqnarray}
with $L_x=\ln(x^2\mu_X^2)=\ln(x^2 e^{2\gamma_E}/4)$.

The spectral density is obtained by calculating the discontinuity
of
\begin{equation}
2\pi^{2-\eps}\int_0^\infty\left(\frac{px}2\right)^{\eps-1}
  J_{1-\eps}(px)(x^2)^{-a}x^{3-2\eps}dx
\end{equation}
for the cases $a=7$ and $a=7-\eps$, where $J_\lambda(z)$ is the Bessel
function of the first kind~\cite{Groote:2005ay}. One obtains ($s=p^2$)
\begin{eqnarray}\label{spectral}
\rho(s)/\rho_0(s)&=&360\left(1+\frac{\alpha_s}{\pi}
  \left(\frac{617}{35}+3L\right)\right)\nonumber\\&&
  -6\left(1+\frac{\alpha_s}{\pi}\left(\frac{6367}{105}+11L\right)\right)
  \nonumber\\&&
  -4\cdot 6\left(1+\frac{\alpha_s}{\pi}\left(\frac{1746}{35}+9L\right)\right)
\end{eqnarray}
with $L=\ln(\mu^2/s)$ and $\rho_0(s)=s^5/378000(4\pi)^8$.

\section{Discussion}
The first NLO contribution in both the correlator and the spectral density 
comes from the factorizing part of the diagrams, i.e.\ the case
where baryon and meson correlators are multiplied in configuration space and
do not mix by gluon exchange. This contribution dominates the final result but
is probably irrelevant for physics as it does not contain a pentaquark bound
state. The physically relevant mixing contributions are smaller since they
are suppressed both in the number of colours $N_c$ and a factor $4$ coming
from the evaluation of Dirac traces.

The result for the spectral density~(\ref{spectral}) shows that the NLO
corrections to the spectral density are large, where the NLO corrections to
the factorizing parts are somewhat smaller (note, however, that the relevant
scale is not $s$ but rather $s/5$ because of the number of lines in the
correlator). This spoils the conventional momentum space QCD sum rule
analysis. The NLO corrections to the correlator function~(\ref{PiRx2}) in
configuration space are more reasonable. This suggests a sum rule analysis in
configuration space, based on the correlator~(\ref{PiRx2}). However, it is not
clear whether the accuracy of such a sum rule analysis in configuration space
will be sufficient for physical applications.

\section{Conclusion}
We have calculated NLO perturbative QCD corrections for pentaquark current
correlators which turn out to be very large. One has to conclude that the NLO
QCD sum rule analysis of pentaquark states is fraught with difficulties.
This complicates the mass determination of the pentaquark states using a QCD
sum rule analysis. One may have to take recourse to model calculations to
determine the properties of pentaquark states such as the one based on chiral
solitons~\cite{Diakonov:1997mm}. 

\begin{acknowledgments}
A.A.P.\ was supported in part by the DFG, Germany, under contract
436~RUS~17/5/05 and by the Russian Fund for Basic Research (05-01-00992a).
S.G.\ acknowledges support by the DFG as a guest scientist in Mainz supported
by the DFG under contract 436~EST~17/1/04, by the Estonian target financed
project No.~0182647s04, and by the grant No.~6216 given by the Estonian
Science Foundation.
\end{acknowledgments}

\end{document}